# Multi-dimensional Separation Prior to Mass Spectrometry:

# Getting Closer to the Bottom of the Iceberg

Manuel MAYR(1) & Thierry RABILLOUD (2,3,4)

1-King's British Heart Foundation Centre, King's College London, London UK

2-Pro-MD team, UMR CNRS 5249, Laboratoire de Chimie et Biologie des Métaux, UMR CNRS-CEA-UJF, Grenoble

3-Pro-MD team, Laboratoire de Chimie et Biologie des Métaux, UMR CNRS-CEA-UJF, Université Joseph Fourier, Grenoble

4-Pro-MD team, CEA Grenoble, iRTSV/LCBM, Laboratoire de Chimie et Biologie des Métaux, UMR CNRS-CEA-UJF, Grenoble

Abstract:
The combination of three separation steps prior to mass spectrometry, respectively SDS electrophoresis of proteins, isoelectric focusing of tryptic peptides then reverse phase chromatography of peptides, deeply increases the coverage of the proteome [1], while keeping the dispersion of the data within a reasonable range. It is thus anticipated that this combination of separations will provide a further step forward in the analysis of complex proteomes

Keywords
Proteome; peptide isoelectric focusing; mass spectrometry; SDS electrophoresis

Undersampling of the proteome remains one of the major issues in proteomics. It is one of the main limitations of 2D gel-based proteomics but equally applies to shotgun proteomics. Even though the latest mass spectrometers perform thousands of MS/MS events per run, only a relatively small percentage of the resulting spectra (usually in the order of 10-20%) are actually assigned to peptides [2]. Incomplete databases, low abundance of the precursor, post-translational modifications, non-tryptic cleavages, interferences by co-eluting peptides or contaminants may all hamper correct peptide identifications. To overcome undersampling, various fractionation schemes have been proposed prior to the final peptide separation step by reverse phase liquid chromatography. Increasing the number of separation steps, however, has two major drawbacks: First, it increases the mass spectrometry time required to analyze the proteome of interest. Second, fractionation introduces variation in the proteomics workflow, as documented recently for several fractionation schemes [3]. Whereas reduced throughput is commonly regarded as a reasonable trade-off to increase proteome coverage, additional variability is problematic. Inevitably, quantitative accuracy is compromised and information gets lost if the technical variability exceeds the biological differences in the samples. Thus, it remains difficult to find the right balance between sensitivity and reproducibility.

Two-dimensional fractionations (e.g. SDS PAGE followed by reverse phase) are widely used but not satisfactory with regards to sensitivity and proteome coverage. It is thus tempting to introduce a third separation step, as recently proposed for protein-based separations [4]. In this issue, Urlaub and colleagues [1] report that adding a third dimension, namely isoelectric focusing (IEF) of tryptic peptides after SDS-PAGE separation and in-gel digestion of proteins offers improved resolution by reverse phase liquid chromatography and thus a major increase in proteome coverage, but adds comparably little technical variability. Increasing the LC gradient time in the gel-LC-MS/MS experiment resulted in fewer new protein identifications than adding an additional dimension by IEF. It appears that with the method proposed by Urlaub *et al* [1] the separation of peptides by IEF has improved peptide resolution to such an extent that more of the low abundant peptides get sampled by the mass spectrometer. Unlike the conventional gel-LC-MS/MS approach, peptides belonging to a particular protein are not just contained within neighbouring gel slices but spread out according to their isoelectric points across the entire experiment. This could also be advantageous for posttranslational modifications that alter the isoelectric point by allowing a better separation of the modified peptides from the non-modified counterparts.

In summary, the three-dimensional workflow for shotgun proteomics established by Urlaub *et al* takes proteomics a step closer to the "bottom of the iceberg" of the cellular proteome.